\documentclass[aps,twocolumn]{revtex4}

\begin{document}

\title{Prospects for application of ultracold Sr$_2$ molecules in precision measurements}

\author{S. Kotochigova$^1$, T. Zelevinsky$^2$, and Jun Ye$^3$}

\address{$^{1}$ Department of Physics, Temple University, Philadelphia,
PA 19122, USA\\
$^2$ Department of Physics, Columbia University, 538 West 120th Street,
New York, NY 10027-5235, USA \\
$^3$ JILA, National Institute of Standards and Technology and University
of Colorado, Boulder, CO 80309-0440, USA }

\begin{abstract}
Precision measurements with ultracold molecules require development of robust and sensitive techniques
to produce and interrogate the molecules.  With this goal, we theoretically analyze
factors that affect frequency measurements between rovibrational levels of
the Sr$_2$ molecule in the electronic ground state. This measurement can be used to constrain the possible time variation of the proton-electron mass ratio. Sr$_2$ is expected to be a strong candidate
for achieving high precision due to the spinless nature and ease of cooling and perturbation-free
trapping of Sr \cite{Zelevinsky2008}.  The analysis includes calculations of
two-photon transition dipole moments between deeply and weakly bound
vibrational levels, lifetimes of intermediate excited states, and
Stark shifts of the vibrational levels by the optical lattice field, including
possibilities of Stark-cancellation trapping.
\end{abstract}

\maketitle

\section{Introduction}

Molecular systems possess a variety of properties that open the possibility
to use them for high-precision measurements.  The richness
of their electronic, vibrational, and rotational spectra
provides a series of precision benchmarks from the radio frequencies to the
visible spectrum. One of the earliest applications of molecules in spectroscopy involved high accuracy
frequency standards. The first molecular clock based on microwave transitions in ammonia
was built in 1949 \cite{Townes}.  Other molecular frequency standards provided
secondary, relative frequency references to enhance measurement precision \cite{Ye},
and played an important role in measurements of fundamental constants such as the speed of light
\cite{Evenson} and possibly the proton to electron mass ratio \cite{Koelemeij}.

Furthermore, molecules are valuable for tests of fundamental physics.
Diatomic molecules have been proposed for use in parity violation studies
\cite{Kozlov,Flambaum,DeMille} due to the enhanced sensitivity of
rovibrational spectra to nuclear effects, as well as in measurements of
the electron electric dipole moment
\cite{Kozlov1,Hudson} that can benefit from very large internal electric fields
of polar molecules.

Recent developments in precision measurement techniques, as well as in cooling
and trapping, have stimulated an interest in the
application of molecular spectroscopy to studies of
possible time variation of fundamental constants
\cite{Flambaum3}. In particular,
research has focused on the variation of the fine structure constant
$\alpha$ \cite{EHudson} and  the proton to electron mass ratio $\mu\equiv m_p/m_e$
\cite{Flambaum2,Chardonnet,Zelevinsky2008,DeMille2008,Flambaum1}.
The values of these constants can drift monotonically, or vary
periodically with the distance from the earth to the sun in case of the existence
of gravitational coupling.  While state-of-the art optical atomic clocks
have set the most stringent limits on both types of $\alpha$ variations \cite{Rosenband2008,Blatt2008},
atoms generally lack transitions that can reveal relative $\mu$ variations ($\Delta\mu/\mu$) in a
model-independent way.  On the other hand,
molecules exhibit complex structure and dynamics due to the combination of their
electronic interaction and the
vibrations and rotations of their constituent nuclei.
For example, if $m_p$ drifts relative to $m_e$, the effect on the weakly bound
vibrational levels near dissociation is expected to be much smaller than on the more deeply bound
levels at intermediate nuclear distances (e.g. see Fig. 2 (b) in Ref. \cite{Zelevinsky2008}).
This can allow accurate spectroscopic determinations of $\Delta\mu/\mu$ by using the least
sensitive levels as frequency anchors.  Moreover, two-color optical Raman spectroscopy
of vibrational energy spacings within a single electronic potential takes advantage of the
entire molecular potential depth in order to minimize the relative measurement error.

\begin{figure}
\vspace*{9.5cm}
\includegraphics{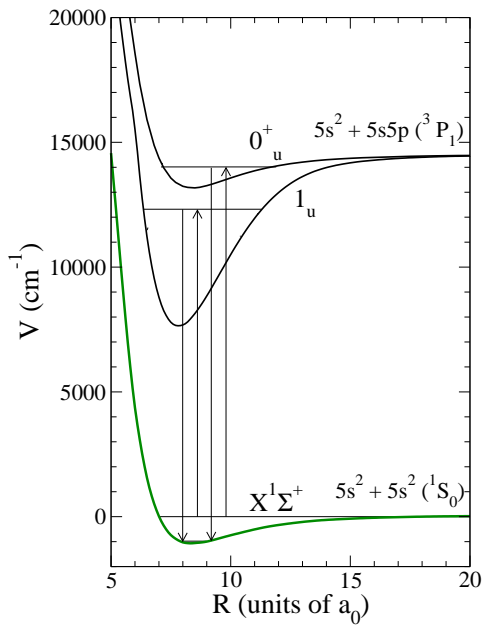}
\caption{Sr$_2$ potential energy diagram that shows vibrational levels of the
1(0$_u^+$) and 1(1$_u$) potentials that can be optically excited from vibrational levels of the ground
state X$^1\Sigma^+$.  The energy zero is at the dissociation limit of the ground state.}
\label{schema}
\end{figure}

For a given spectral line at frequency $\nu$, the systematic line shift $\Delta\nu$ results from
$\Delta\mu$ through the proportionality constant $\kappa$,
\begin{equation}
\frac{\Delta\mu}{\mu}=\kappa\frac{\Delta\nu}{\nu}.
\label{DeltaMuDeltaNu}
\end{equation}
For direct spectroscopy of molecular vibrational energy levels, $\kappa\sim1$.
To minimize the fractional uncertainty $\delta\mu/\mu$ on the measurement of
$(\Delta\mu\pm\delta\mu)/\mu$, we must minimize $\kappa(\delta\nu/\nu)$, where $\delta\nu$ is the measurement uncertainty of the absolute frequency $\nu$.  Since in precision spectroscopy experiments the limitations are typically on $\delta\nu$ rather than the relative uncertainty $\delta\nu/\nu$, and since
\begin{equation}
\frac{\delta\mu}{\mu}=\kappa\frac{\delta\nu}{\nu}=\left(\frac{d\nu}{d\ln\mu}\right)^{-1}\delta\nu,
\label{ErrorMin}
\end{equation}
the quantity $\nu/\kappa=d\nu/d\ln\mu$ must be maximized.  This implies that the optimal
frequency intervals for the $\mu$-variation-sensitive molecular clock are
those with the largest absolute frequency
shifts ($\Delta\nu$) due to a given fractional change $\Delta\mu/\mu$ (for Sr$_2$, this is approximately
270 cm$^{-1}$ for a unity change in the mass ratio). In some proposed schemes \cite{DeMille2008,Flambaum3}, near-degeneracies of vibrational levels from different electronic potentials permit a frequency measurement of $\nu$ to be carried out in the microwave domain, resulting in a small $\delta\nu$. The effective $\kappa$ is small, and thus the corresponding quantity $\nu/\kappa$ is reasonably large. In the proposed optical Raman measurement \cite{Zelevinsky2008}, however, the sensitivity enhancement comes from the cumulative shift effect over the depth of the molecular potential.  While $\kappa$ is on the order of unity here, $\nu$ is large, resulting in a large $\nu/\kappa$. Furthermore, if the least sensitive energy gap is measured concurrently with the most sensitive one, it can serve as a reference and thus remove
any frequency drift of an intermediate (atomic) clock used in the measurement.
Thus, in our previous work \cite{Zelevinsky2008} we have proposed measuring the intervals between
the Sr$_2$ X$^1\Sigma^+$ (see Fig. \ref{schema}) vibrational
levels $v=27$ (in the middle of the potential well) and $v=-3$ (near
dissociation, counting from the top), as well as between $v=27$ and $v=0$.  These are complementary
schemes, with opposite dependences on $\Delta\mu/\mu$.  The difference between these two frequency intervals,
normalized by their sum, doubles the sensitivity of the measurement, while eliminating any
drift of an intermediate frequency reference used to stabilize the Raman lasers.
Finally, all-optical molecular spectroscopy in the tight confinement (Lamb-Dicke) regime
in an optical lattice can be Doppler and recoil free, leading to small frequency
uncertainties $\delta\nu$.

Ultracold homonuclear molecules in optical lattices are particularly good candidates
for high precision measurements such as those of $\Delta\mu/\mu$.  Their vibrational
levels have long natural lifetimes that are insensitive to black body radiation,
and the molecules do not experience long-range interactions when isolated in
optical lattice sites.  In particular, Sr$_2$ dimers are promising for
precise molecular metrology \cite{Zelevinsky2008}.  Sr atoms can be directly
laser cooled to temperatures of $\sim1$ $\mu$K \cite{KatoriPRL1999,LoftusPRA2004}, and can be trapped
in Stark-cancellation optical lattices that eliminate spectral shifts and
broadening \cite{YeScience2008}.  Optical atomic clocks based on forbidden
electronic transitions in $^{87}$Sr have shown that systematic frequency uncertainties
can be reduced to the $10^{-16}$ level \cite{LudlowScience2008} in the Lamb-Dicke regime
of a Stark-cancellation lattice.  Moreover, work on $^{88}$Sr
narrow line $^1S_0-^3P_1$ photoassociation \cite{Zelevinsky06} has shown high experimental efficiency and resolution,
and a potential for excellent agreement with theoretical calculations.  The comparisons between
experiments and theory are largely facilitated by the spinless nature of the $^{88}$Sr ground
state ($^1S_0$, total electronic angular momentum $J=0$, nuclear spin $I=0$).  The electronic ground state of Sr$_2$ thus corresponds
to a single molecular potential (X$^1\Sigma^+$ in Fig. \ref{schema}) with non-degenerate vibrational levels with
the exception of rotational duplicity.  This simple ground state potential
should allow highly efficient spectroscopic addressing of molecular vibrational levels.  Finally,
the laser-accessible metastable molecular potentials (corresponding to the $^1S_0+^3P_1$ limit) are expected to have very large Franck-Condon overlaps with the electronic ground state \cite{Zelevinsky06,Zelevinsky2008}, further enhancing
the atom-molecule conversion efficiency and the Raman molecular vibrational transition strengths.

The proposed mechanism for inducing the vibrational transitions is optical Raman spectroscopy \cite{Ni2008}, using
vibrational levels of metastable Sr$_2$ as intermediate states.
The Sr$_2$ ground state electronic potential is 30 THz deep \cite{Czuchaj03}, and the
Raman lasers can be stabilized to $<0.1$ Hz relative to each other \cite{LudlowLaser07}.
Assuming availability of an optical frequency standard \cite{LudlowScience2008},
a power broadened linewidth of 10 Hz, and a signal to noise ratio of
100, the fractional frequency precision (or, equivalently, the precision
of the $\Delta\mu/\mu$ test) approaches
$\sim5\times10^{-15}/\sqrt{\tau}$, where $\tau$ is the averaging time in seconds
\cite{Zelevinsky2008}.  Achieving this precision would lead to an important contribution
to the $\mu$ variation data.  Presently, a combination of atomic clock data
constrains $\Delta\mu/\mu$ to $\sim4\times10^{-16}$/year \cite{Rosenband2008}, and the
evaluation of astronomical NH$_3$ spectra constrains $\Delta\mu/\mu$ to
$\sim3\times10^{-16}$/year \cite{Flambaum2}.  The interpretation of the atomic clock data
is dependent on theoretical modeling, since hyperfine transitions depend
on both $\alpha$ and $\mu$.  The ammonia result is less model-dependent,
but relies on poorly controllable cosmological observations, and
disagrees with the cosmological H$_2$ measurements of $\Delta\mu/\mu$
that indicate non-zero mass ratio variation at the $10^{-15}$/year level \cite{Reinhold06}.
Ultracold molecule-based clock for precision measurements would
provide a model-independent $\mu$-variation-sensitive system with small, controllable
experimental uncertainties.

In this paper we present a detailed analysis of various factors affecting
frequency measurements between vibrational levels in the ground
state of Sr$_2$ in an optical lattice.
We evaluate transition dipole moments between vibrational levels of the ground and
metastable states, for determination of the optimal pathways for Raman
transitions.  In addition, we analyze dynamic polarizabilities of the
ground state vibrational levels in order to identify Stark-cancellation
optical lattice frequencies for vibrational transitions.  Finally, we
estimate the natural linewidths of the intermediate metastable vibrational
levels, in order to obtain realistic estimates of spontaneous scattering
rates of the optical lattice and Raman spectroscopy photons, and thus of
trap losses \cite{Zelevinsky2008}.  The electronic potentials and their properties
used in the calculations of the vibrationally averaged dipole moments, polarizabilities, and
linewidths are presented in Ref. \cite{Kotochigova2008}.

\section{Dipole moments and Raman transitions}
\label{Sec:DMs}

For the production of Sr$_2$ dimers in a particular rovibrational level
of the ground state potential (X$^1\Sigma^+$), it is necessary to study
the transition dipole moments and transition probabilities between
vibrational levels of the X$^1\Sigma^+$ and the metastable 1(0$^+_u$)
and 1(1$_u$) potentials (dissociating to the $^1S_0+^3P_1$ limit) shown
in Fig.~\ref{schema}.  The numerical labels ($\Omega$) of the \textit{ungerade}
potentials $0_u^+$ and $1_u$ correspond to the total atomic angular
momentum projection onto the internuclear axis of 0 and 1, respectively.
We choose to work with the metastable intermediate states because of
the high spectroscopic resolution and low scattering rates that are
possible with narrow lines, and due
to the large estimated Franck-Condon overlaps with vibrational levels of
X$^1\Sigma^+$ \cite{Zelevinsky06,Zelevinsky2008}.  Preliminary estimates of
the transition strengths between weakly bound metastable and ground state
levels were obtained in previous work \cite{Zelevinsky06}.  However,
for precise molecular metrology, more deeply bound vibrational levels
must be optically coupled via a Raman scheme.  There is no information on these optical
transitions in the literature.

The transition dipole moment between vibrational
level $v'J'M'$ of an excited-state $\Omega'=$0,1 \textit{ungerade} potential
and vibrational level $vJM$ of the ground-state $\Omega=0$ (X$^1\Sigma^+$) \textit{gerade}
potential is given by
\begin{eqnarray}
   \lefteqn{\langle\Omega'\, v'J'M' \, | \, d(R)C_{1\epsilon}(\hat R) \, | \,
                                     {\rm X} ^1\Sigma^+\, vJM \rangle
   }\\ \nonumber
   && =\langle\Omega'\, v'J' \, || \, d(R) \, || \, {\rm X} ^1\Sigma^+\, vJ  \rangle
              \nonumber\\
   &&
      \quad  \quad   \times \sqrt{(2J'+1)(2J+1)}\
           (-1)^{\epsilon-\Omega'+M} \nonumber \\
   &&
    \quad  \quad  \quad  \quad  \times
         \left( \begin{array}{ccc}
                          1 &J & J' \\
                          -\epsilon & -M & M'
                     \end{array}\right)
              \left( \begin{array}{ccc}
                          1 & J & J' \\
                          -\Omega' & 0 & \Omega'
                     \end{array}\right) \, , \nonumber
\end{eqnarray}
where $\hat{R}$ is the orientation of the interatomic axis,
$C_{1\epsilon}(\hat R)$ is a spherical harmonic, $\vec{\epsilon}$ is the
polarization of the laser field, and $M$, $M'$ are projections along a laboratory
fixed coordinate axis of the total angular momenta $J$, $J'$.
The vibrationally averaged or reduced
matrix element $\langle\Omega'\, v'J'|| d(R) \, || \, {\rm X}^1\Sigma^+\,
vJ\rangle$ is a radial integral  over the $R$-dependent relativistic
electronic dipole moment $d(R)$ and rovibrational wave functions of the
electronic state $|\Omega'\rangle$ and ground electronic state
$|{\rm X}^1\Sigma^+\rangle$.

\begin{figure}
\vspace*{8cm}
\includegraphics{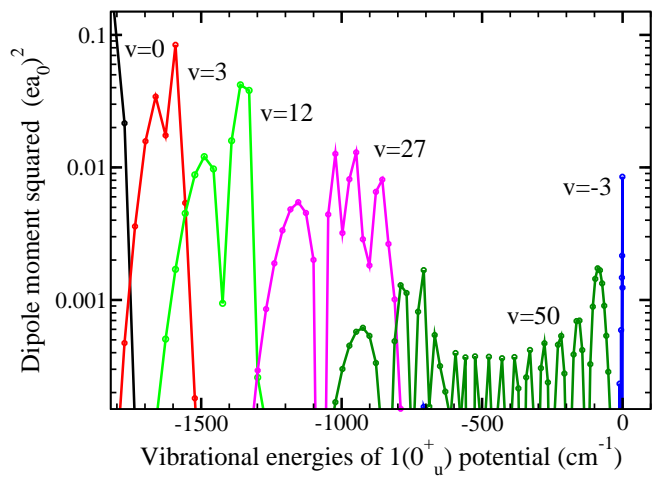}
\caption{(Color online) The square of the vibrationally averaged dipole moments
for transitions from selected vibrational levels (all with $J=0$) of the
X$^1\Sigma^+$ electronic ground state as a function of the
energy of vibrational levels (with $J'=1$) of the 1(0$^+_u$) potential of Sr$_2$.
Each curve corresponds to a single rovibrational level of the X state.
Zero energy corresponds to the $^1S_0+^3P_1$ dissociation limit.
}
\label{FC_0+}
\end{figure}

\begin{figure}
\vspace*{8cm}
\includegraphics{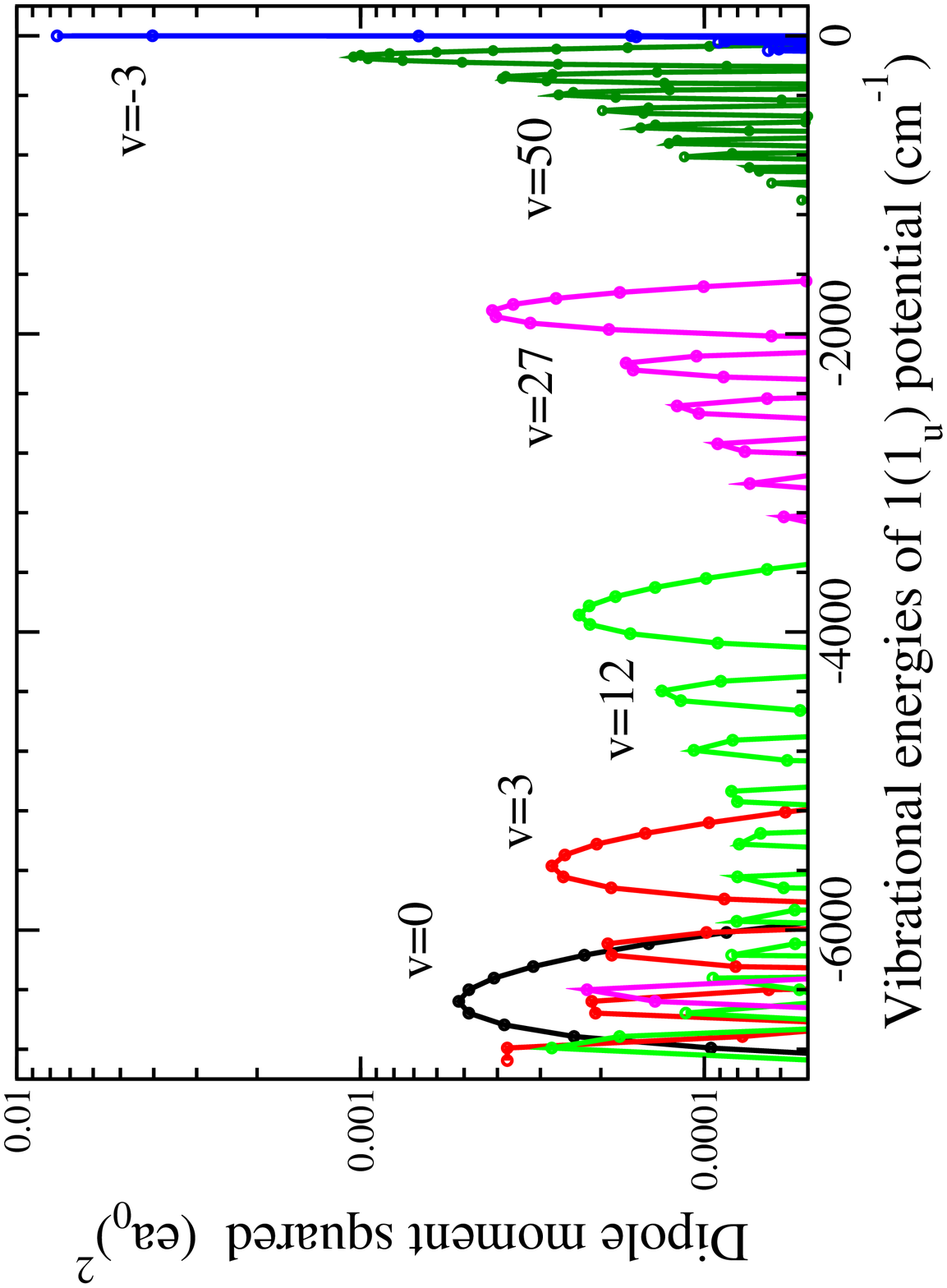}
\caption{(Color online) The square of the vibrationally averaged dipole
moments for transitions from selected vibrational levels (all with $J=0$) of the X$^1\Sigma^+$
electronic ground state as a function of the energy of vibrational levels
(with $J'=1$) of the 1(1$_u$) potential of Sr$_2$.  Each curve
corresponds to a single rovibrational level of the X state.  Zero energy
corresponds to the $^1S_0+^3P_1$ dissociation limit.}
\label{FC_1}
\end{figure}

\begin{figure}
\vspace*{8cm}
\includegraphics{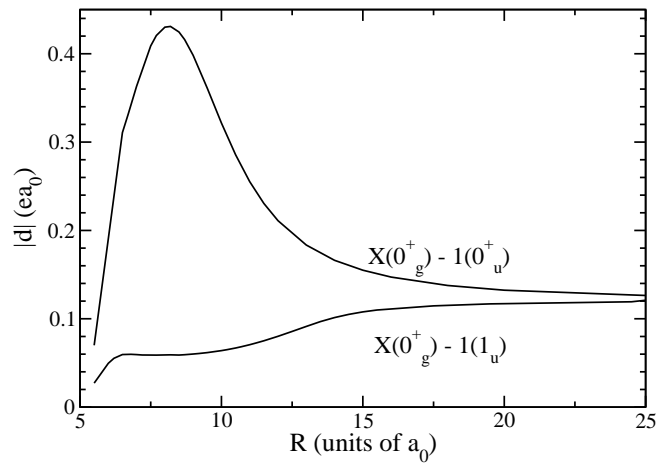}
\caption{Absolute values of electronic transition dipole moments from the X$^{1}\Sigma^+$ ground state
to the metastable states 1(0$^+_u$,1$_u$) of Sr$_2$. Adapted from \cite{Kotochigova2008}.}
\label{dm_triplet}
\end{figure}

We study transitions to both metastable potentials dissociating to $^1S_0+^3P_1$
(Fig.~\ref{schema}). Figures \ref{FC_0+} and \ref{FC_1} show the squares of the vibrationally
averaged dipole moments as a function of vibrational energies of the
potentials 1(0$^+_u$) and 1(1$_u$), respectively, in atomic units $(ea_0)^2$, where $e$ is the electron
charge and $a_0=0.053$ nm is the Bohr radius.  A comparison of the
two figures shows that the squares of the dipole moments for 1(0$^+_u$)
are two orders of magnitude larger than for 1(1$_u$) for deeply bound
levels.  This is due to the difference in strengths of the electronic
transition dipole moments at small separations (shown in Fig.~\ref{dm_triplet})
and the similarity in shape of the X$^1\Sigma^+$ and 1(0$^+_u$) potentials.
This similarity also explains the localized character of
Franck-Condon factors for a given vibrational level of the X$^1\Sigma^+$ state (see Fig. 2).
However, for weakly bound levels with binding energies less than 500 cm$^{-1}$ the squared dipole
moments for the two potentials have similar magnitudes.
The electronic dipole moments near the outer turning points have similar magnitudes,
and the 1(0$^+_u$) and 1(1$_u$) potentials are more similar in shape.
In our analysis of Raman transition rates we will focus on the
1(0$^+_u$) potential since it has favorable dipole moments with the
ground state.

\begin{figure}
\vspace*{8cm}
\includegraphics{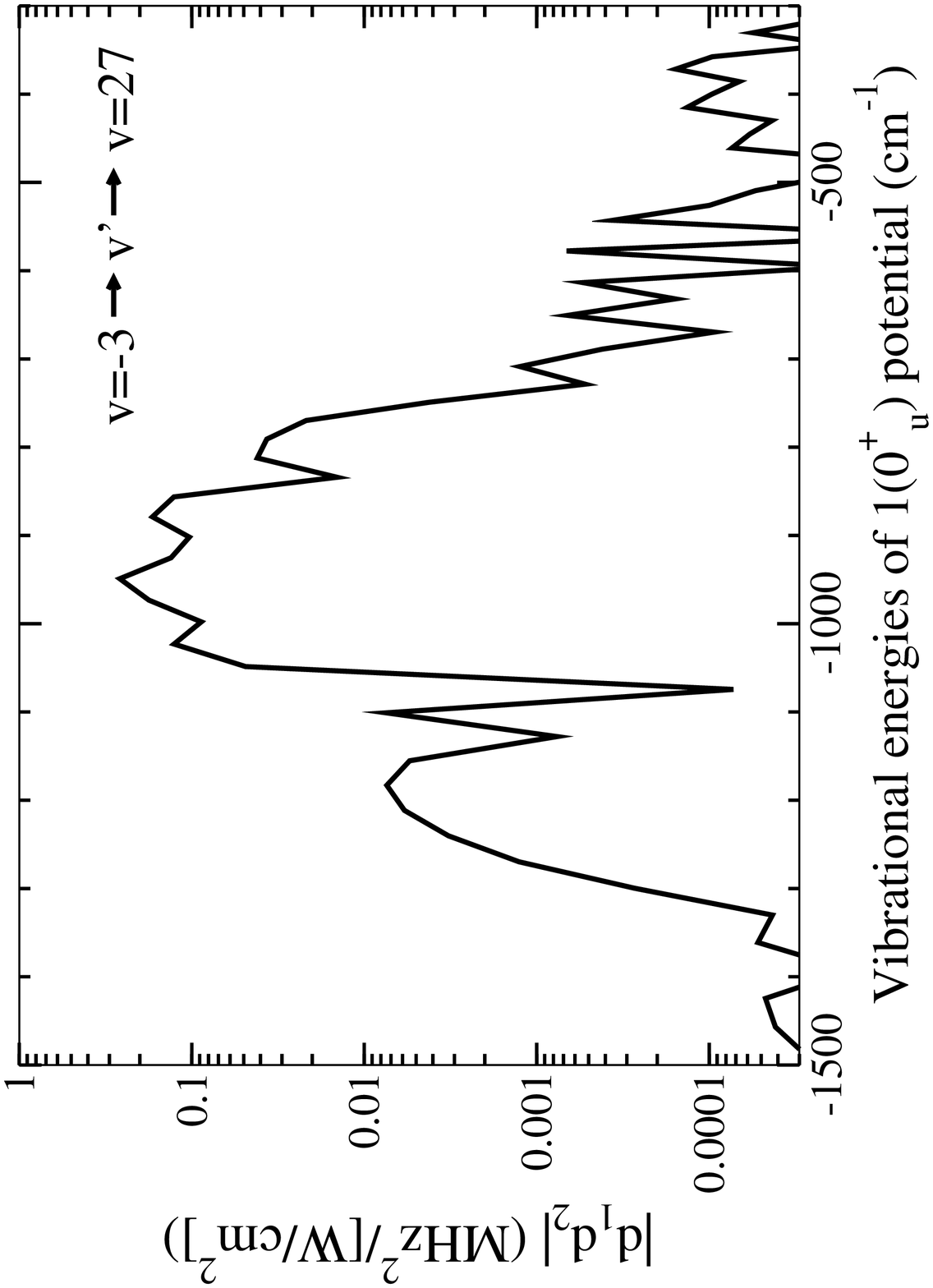}
\caption{Two-photon vibrationally averaged dipole moments
for the Raman transitions as a function of the energies of $v', J'=1$ rovibrational
levels of the 1(0$^+_u$) intermediate state.  The pathway that was proposed in
Ref.~\cite{Zelevinsky2008} is shown. It starts from the $v=-3, J=0$
rovibrational level and ends at the $v=27, J=0$
level of the X$^1\Sigma^+$ ground state.
The energy zero corresponds to the $^1S_0+^3P_1$ dissociation limit.}
\label{Raman-3}
\end{figure}

A Raman transition is the  process by which we transfer population
from an initially occupied vibrational level to a final level via an
intermediate state.  In this case, both the initial and final levels are
$J=0$ rovibrational levels of the X$^1\Sigma^+$ potential while the intermediate level is
a $J'=1$ rovibrational level of 1(0$^+_u$).  The
Rabi matrix element for this two-photon process is proportional to the product
of the dipole moments of both transitions.
We chose the weakly bound $v=-3$ level of the X$^1\Sigma^+$ potential as the initial Raman state, since it is expected to be most efficiently populated by photoassociation \cite{Zelevinsky06}.  Figure \ref{Raman-3} shows the dipole moment products for the $v=-3$ initial state and  the $v=27$ final state as a function of the vibrational energy of the 1(0$^+_u$) potential. This pathway was selected in Ref.~\cite{Zelevinsky2008} for the precision measurement of possible variations of
the proton to electron mass ratio.

\section{Stark shifts in an optical lattice}

Trapping ultracold molecules in optical lattices via ac Stark shifts is key for optimized
control and precision \cite{Zelevinsky2008}.  Optical lattice traps benefit both the
reduction of systematic effects, and the achievement of high molecule densities on the order of
$10^{12}$/cm$^3$ \cite{Zelevinsky06}.  The Stark-cancellation, or \textit{magic frequency},
technique \cite{YeScience2008} has enabled state-of-the-art neutral atom clocks \cite{Blatt2008}.
This approach relies on a suitable crossing of dynamic polarizabilities of the two probed
states at a particular lattice wavelength.  This ensures a zero differential Stark shift,
and a suppression of the inhomogeneous Stark broadening.  Analogously, Stark-cancellation
optical lattice frequencies can be sought for specific pairs of the Sr$_2$ vibrational levels.
For example, tuning the lattice frequency near a narrow resonance associated with an optical
transition from a vibrational level of X$^1\Sigma^+$ to another one of 1(1$_u$) can help achieve matching polarizabilities of a vibrational level pair
of X$^1\Sigma^+$.

The ac Stark shifts of the vibrational levels of illuminated ground state molecules
are determined by the dynamic polarizability $\alpha(h\nu,\vec{\epsilon})$, which is
a function of the radiation frequency $\nu$ and polarization $\vec{\epsilon}$
($h$ is the Planck constant).
If the Sr$_2$ molecule is in the
ground state X$^1\Sigma^+$, its dynamic polarizability in SI units is
expressed in terms of the dipole coupling to the rovibrational levels of the
excited potentials as
\begin{eqnarray}
  \alpha(h\nu,\vec{\epsilon}) = \frac{1}{\epsilon_0c}
\label{polariz}
   \sum_{f} \frac{(E_f - ih\gamma_f/2 - E_i)}{(E_f - ih\gamma_f/2 - E_i)^2 - (h\nu)^2}\\ \nonumber
     \times |\langle f|d(R) \,\hat{R}\cdot \vec{\epsilon}|i\rangle|^2,
\end{eqnarray}
where $c$ is the speed of light, $\epsilon_0$ is the electric constant,
$\langle f|d|i \rangle$
are $R$-dependent electric dipole moments (containing both radial and angular
contributions), and $i$ and $f$ denote
the initial $|vJM\rangle$ and intermediate $|v'J'M'\rangle$
rovibrational wave functions of  the X$^1\Sigma^+$  and $\Omega'$ electronic states,
respectively.  The energy $E_i$ is the rovibrational energy of the X$^1\Sigma^+$
state and $E_f$ is the rovibrational energy of the
$\Omega'$ state.  Finally, the line widths $\gamma_f$
describe the spontaneous and other decay and loss mechanisms.
Equation~(\ref{polariz}) is a sum over dipole transitions to the rovibrational levels of excited
potentials; this includes contributions from the continuum of the $\Omega'$ states as well.
This sum can be truncated if transitions have negligible
electric dipole moments or large detunings. In the case of Sr$_2$,
the sum only includes contributions from potentials with
$\Omega'=0,1$ and \textit{ungerade} symmetry.  The potentials and transition dipole moments
are calculated in Ref. \cite{Kotochigova2008}.

The frequency of the optical lattice is in the near infrared range, where
the laser can be tuned close to a transition from
a ground state vibrational level to a vibrational level of the
1(1$_u$) potential (Fig. \ref{schema}).
We employ two approaches to finding either vibrational levels with the same Stark shift, or
the Stark-cancellation light frequency for a given pair of levels.
The first approach is to analyze the dependence of the baseline molecular
polarizability on the vibrational quantum number for a fixed laser
frequency and look for non-monotonic behavior, thus allowing certain pairs of
vibrational levels to have the same Stark shifts.  The second approach
is to rely on sharp polarizability resonances arising from presence of vibrational levels
in the metastable potentials as shown in Fig. \ref{schema}.  While the latter method
allows significantly more flexibility, care must be taken
to ensure low incoherent photon scattering rates \cite{Zelevinsky2008}.

\begin{figure}
\vspace*{8cm}
\includegraphics{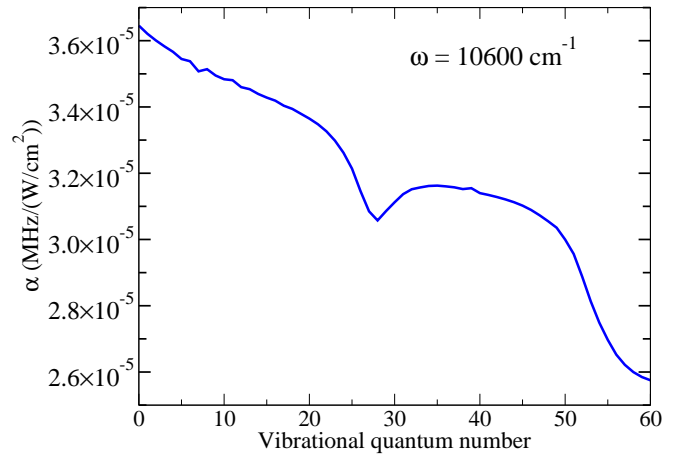}
\caption{
Real part of the dynamic polarizability $\alpha$ of the $J=$0 levels
of the X$^1\Sigma^+$ ground state of Sr$_2$ as a function of vibrational
quantum number, for infrared laser frequency of 10600 cm$^{-1}$.}
\label{infra_red}
\end{figure}

For simultaneous trapping of Sr$_2$ molecules in different vibrational levels
of the X$^1\Sigma^+$ potential in a Stark-cancellation regime, we can use the
first approach and search for
vibrational levels of X$^1\Sigma^+$ that have equal polarizabilities.
Figure \ref{infra_red} shows the dynamic polarizability at the infrared
laser frequency of 10600 cm$^{-1}$ as a function of the vibrational quantum
number of the X$^1\Sigma^+$ potential for the rotational level $J$=0.
The polarizability tends to decrease with increasing vibrational quantum number;
exceptions occur for $v > 27$ vibrational levels. Analysis of the individual contributions to
the polarizability shows that the dip near $v = 27$ is due to an avoided crossing between relativistic
$\Omega = 1$ excited potentials, which coincides with the outer turning point of the $v = 27$
vibrational level of the ground state potential. As a result, we can find matching polarizabilities for the
higher vibrational levels. However, there is a possibility that this dip is an artifact of our method of calculation.
We assumed that the vibrational motion is purely adiabatic, which might not be accurate for this
avoided crossing. It would require a multichannel calculation of the vibrational levels to test our results.

\begin{figure}
\vspace*{8cm}
\includegraphics{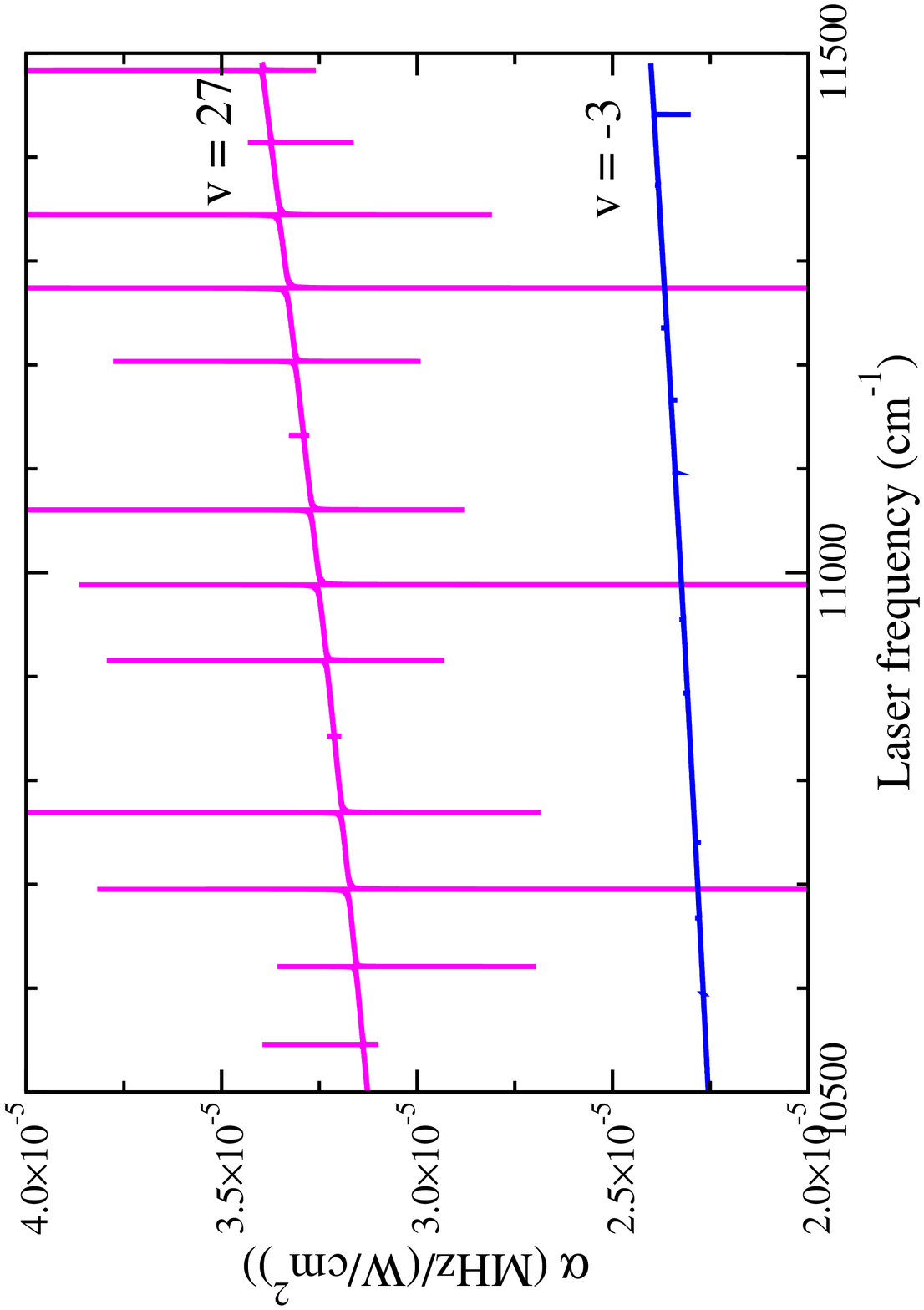}
\caption{(Color online) Real part of the dynamic polarizability of  the $J=0$ vibrational
levels of the Sr$_2$ X$^1\Sigma^+_g$ state as a function of laser
frequency in the range of 10500-11500 cm$^{-1}$.  For clarity, only two
selected vibrational levels of the X state are shown, $v=-3$ and $v=27$.
The polarizabilities are evaluated at 0.1 cm$^{-1}$ intervals.
The polarizabilities of $J=0$ rotational levels are independent of
the laser polarization.}
\label{polar}
\end{figure}

\begin{figure}
\vspace*{8cm}
\includegraphics{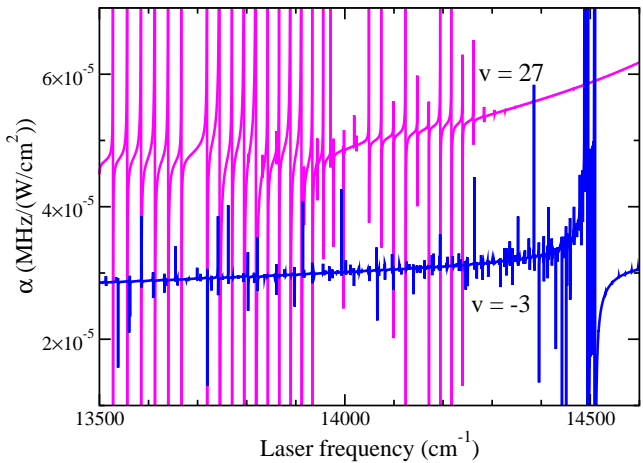}
\caption{(Color online) Real part of the dynamic polarizability of  the $J=0$ vibrational
levels of the Sr$_2$ X$^1\Sigma^+_g$ state as a function of laser
frequency in the range of 13500-14600 cm$^{-1}$.
For clarity, only two selected vibrational levels of the X state are shown, $v=-3$ and $v=27$.
The polarizabilities are evaluated at 0.1 cm$^{-1}$ intervals.  The polarizabilities
of $J=0$ rotational levels are independent of the laser polarization.}
\label{polar1}
\end{figure}

The second approach to matching the polarizabilities of different vibrational levels makes use of
resonances. The calculated molecular polarizabilities of the Sr$_2$ ground state vibrational
levels of interest are shown in Fig.~\ref{polar} as a function of laser frequency in the vicinity of
910 nm, which is near the Stark-cancellation wavelength for the $^1S_0-^3P_1$ transition for atomic $^{88}$Sr.
The overall polarizabilities are determined by the baseline value of $\alpha\sim$2-4$\times10^{-5}$ MHz/(W/cm$^2$)
arising from far-detuned dipole-allowed transitions to states dissociating to the $^1S_0-^1P_1$  atomic limit,
as well as by the resonant structure due to narrow
vibrational levels of the 1(1$_u$) potential dissociating to the $^1S_0-^3P_1$ limit. Figure \ref{polar}
shows three resonances in the range of 10700 cm$^{-1}$ to 11300 cm$^{-1}$ with matching polarizabilities
for the $v= -3$ and $v= 27$ vibrational levels.

The profiles of these resonances are determined by Franck-Condon
factors between the vibrational levels of the X$^1\Sigma^+$ and 1(1$_u$) potentials.
The natural linewidths of the 1(1$_u$) levels
are below $10^{-6}$ cm$^{-1}$,  which is negligible on the frequency scale
of Fig.~\ref{polar}.   A blowup of one of the resonances in Fig.~\ref{polar} near the frequency
of 10988 cm$^{-1}$ is shown in our previous publication \cite{Zelevinsky2008}.
An optical lattice  frequency near this resonance was identified for Stark cancellation of the two ground state vibrational
levels.

Other resonances could be chosen for optical lattice frequencies to facilitate precision measurements, for example,
those in the laser frequency range of 13500-14600 cm$^{-1}$. These resonances arise from transitions
 between vibrational levels of the X$^1\Sigma^+$ and 1(0$^+_u$) potentials.
Figure~\ref{polar1} shows many such resonances with matching polarizabilities for the $v=-3$ and $v=27$
vibrational levels.  Choosing to work near these resonances requires caution regarding the
possibly enhanced scattering of the lattice light.  Previously, we have estimated
the incoherent scattering rates of only $\sim1$/s with 10 kW/cm$^2$ laser intensities
\cite{Zelevinsky2008}, using the calculated
spontaneous decay rates of 1(1$_u$) (Sec. \ref{Sec:Linewidths}) and the transition dipole moments (Sec. \ref{Sec:DMs}).

The exact locations of the resonances shown in Figs.~\ref{polar} and \ref{polar1} must be refined
with the additional input of experimental data.  To date, fewer than ten most weakly bound states
of $1(0_u^+)$ and $1(1_u)$ have been experimentally identified \cite{Zelevinsky06}.
However, the general polarizability trends, the density of vibrational levels, and the relative
line strengths are expected to remain close to our present predictions.

\section{Natural linewidths of excited vibrational levels}
\label{Sec:Linewidths}

Linewidths of the vibrational levels of the $1(0_u^+)$ and $1(1_u)$ potentials are
important for estimating incoherent scattering rates of the Raman spectroscopy light
and of the Stark-cancellation optical lattice trap, as well as of near-resonant
Stark shifts \cite{Zelevinsky2008}.
The natural decay rate of the vibrational level $v'J'M'$ of an excited potential $\Omega'$
is given by the Einstein $A$ coefficient in s$^{-1}$,
\begin{eqnarray}
   \lefteqn{A =  \frac{8\pi }{3 h c^3}
        \sum_{\epsilon,vJM}  \omega_{vJ}^3  }
      \label{Acoefficient}  \\
        &&\quad\quad\quad\times | \langle \Omega'\,v' J' M' | d(R) C_{1\epsilon}(\hat{R})
                             | {\rm X}^1\Sigma^+ v J M \rangle|^2,
         \nonumber
\end{eqnarray}
where the sum is over all light polarizations $\epsilon$ and  quantum numbers
$vJM$ of the ground state potential, $v$ can denote
either vibrational states or continuum scattering states, and
the quantities $\omega_{v J}/(2\pi)$ are the transition frequencies from the excited
vibrational level to the ground state rovibrational level $|vJ\rangle$.

\begin{figure}
\vspace*{8cm}
\includegraphics{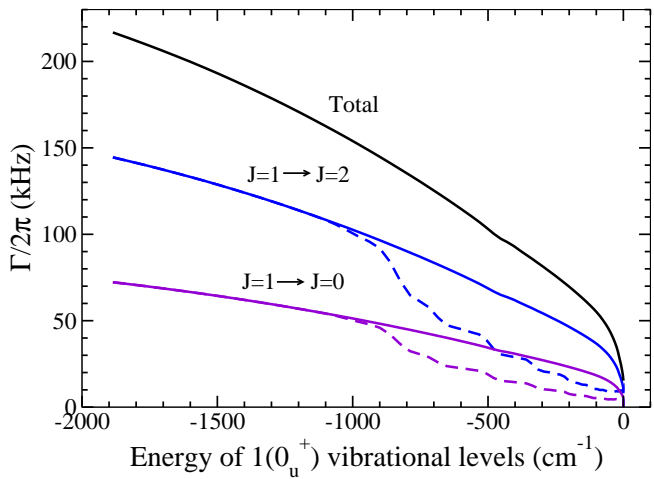}
\caption{(Color online)
Natural linewidths of the 1(0$_u^+$) vibrational levels of Sr$_2$ as a function
of their energy. The individual contributions
from the $(J=1\rightarrow J=0)$ and $(J=1\rightarrow J=2)$ pathways are shown.  Dashed lines
indicate contributions from bound-bound transitions.
The zero energy corresponds to the $^1S_0+^3P_1$  dissociation limit.
}
\label{width0}
\end{figure}

\begin{figure}
\vspace*{8cm}
\includegraphics{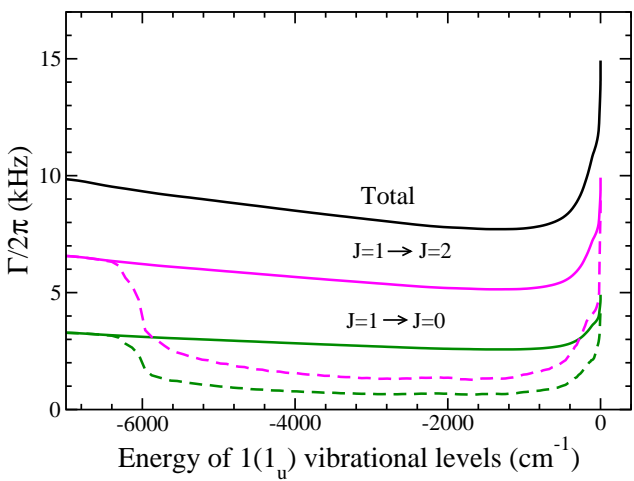}
\caption{(Color online)
Natural linewidths of the 1(1$_u$) vibrational levels of Sr$_2$ as a function
of their energy.  The individual contributions
from the $(J=1\rightarrow J=0)$ and $(J=1\rightarrow J=2)$ pathways are shown.  Dashed lines
indicate contributions from bound-bound transitions.
The zero energy corresponds to the $^1S_0+^3P_1$  dissociation limit.
}
\label{width1}
\end{figure}

Figures \ref{width0} and \ref{width1} show natural linewidths of the vibrational levels of
the excited 1(0$_u^+$) and 1(1$_u$) potentials for all binding energies.
The linewidths contain contributions from transitions between $J=1$
rotational levels of the 1(0$_u^+$) and 1(1$_u$) potentials to $J=0$ and $J=2$
rotational levels of the ground state.  These are the only relevant rotational levels,
since only $J=1$ metastable levels are populated with photoassociation
of the spinless $^{88}$Sr atoms.  The values
and trends of the linewidths are primarily determined by the $R$-dependent
dipole moments between the metastable and ground states plotted in Fig. \ref{dm_triplet}.
For most weakly bound levels, the linewidths approach 15 kHz, in good agreement
with twice the linewidth of the $^3P_1$ atomic limit \cite{Zelevinsky06}.
For deeply bound levels with outer turning points below
14 $a_0$, the transition dipole moments and linewidths are significantly larger
for the 1(0$_u^+$) potential than for 1(1$_u$).

Figures \ref{width0} and \ref{width1} also show the contributions
of bound-bound transitions to the linewidths. For
the deepest and the most weakly bound vibrational levels, nearly all natural decay
is into the bound states.  This is explained by the similar shapes of the
ground state and metastable potentials at small and large internuclear separations.
Previous work \cite{Zelevinsky06} indicated single vibrational channel decay
efficiencies up to 90\% for weakly bound states.
The equilibrium internuclear separations are very similar for the ground state and
both metastable states, which explains the high bound-bound decay efficiencies
for the deeply bound molecules.
At intermediate binding energies, the continuum of the ground state contributes
significantly to the linewidths.

\section{Conclusion}

We have investigated the properties of Sr$_2$ that are relevant to
the development of precision measurement techniques with ultracold molecules.
The experimental and theoretical input from our previous work \cite{Zelevinsky06,Kotochigova2008}
has allowed us to carry out realistic calculations of the two-photon Raman transition
strengths between two targeted vibrational levels in the ground state via an
intermediate metastable state.  We find
the results to favor single-step transfer from the weakly bound to intermediate
vibrational levels of Sr$_2$ in the ground state.  Further, we have explored two
approaches to implementing Stark-cancellation optical lattice traps for Sr$_2$.
We find that the method relying on polarizability resonances should allow
Stark-shift-free trapping with low molecule losses.
The relatively small photon scattering rates are determined by the optical transition
strengths as well as by the total decay rates of the vibrational levels in the metastable electronic state,
which are calculated including the bound-bound and bound-continuum contributions.
The combination of ultracold temperatures, perturbation-free trapping, efficient
optical transfer, and low environmental sensitivity of Sr$_2$ should
allow effective production and precise manipulation and measurements with this
molecule.

We acknowledge financial support from AFOSR, ARO, NIST, and NSF.

\end{document}